\documentclass[twocolumn,aps,pre,superscriptaddress]{revtex4}
\usepackage{color}
\usepackage{graphicx}
\usepackage{amsmath}
\usepackage{amssymb}
\usepackage{amsfonts}
\usepackage{psfrag}

\begin{document}

\title{Yield drag in a two-dimensional foam flow around a circular
       obstacle: \\
Effect of liquid fraction}

\author{Christophe Raufaste}
\thanks{ Address for correspondence: craufast@ujf-grenoble.fr}
\affiliation{Laboratoire de Spectrom\'etrie Physique, BP 87, 38402
St Martin d'H\`eres Cedex, France}
\altaffiliation{UMR 5588 CNRS and
Universit\'e Joseph Fourier.}
\author{Benjamin Dollet}
\thanks{\textit{present address:} Physics of Fluids, University of
Twente, The Netherlands}
\affiliation{Laboratoire de Spectrom\'etrie Physique, BP 87, 38402 St Martin d'H\`eres Cedex, France}
\altaffiliation{UMR 5588 CNRS and Universit\'e Joseph Fourier.}
\author{Simon Cox}
     \affiliation{Institute of Mathematical and Physical Sciences,
University of Wales Aberystwyth, SY23 3BZ, UK}
\author{Yi Jiang}
     \affiliation{Theoretical Division, Los Alamos National Laboratory,
Los Alamos,
NM 87545, USA}
\author{Fran\c cois Graner}
\affiliation{Laboratoire de Spectrom\'etrie Physique, BP 87, 38402
St Martin d'H\`eres Cedex, France}
\altaffiliation{UMR 5588 CNRS and Universit\'e Joseph Fourier}

\date{\today}

\begin{abstract}

We study the two-dimensional flow of foams around a circular
obstacle within a long channel. In experiments, we confine the
foam between liquid and glass surfaces. In simulations, we use a
deterministic software, the Surface Evolver, for bubble details
and a stochastic one, the extended Potts model, for statistics. We
adopt a coherent definition of liquid fraction for all studied
systems. We vary it in both experiments and simulations, and
determine the yield drag of the foam, that is, the 
force exerted on the obstacle by the foam flowing at very low velocity. We find that the yield drag  is linear
over a large range of the ratio of obstacle to bubble size, and
is independent of the channel width over a large range. Decreasing the liquid
fraction, however,  strongly increases the yield
drag; we discuss and interpret this dependence.
\end{abstract}

\maketitle

\section{Introduction}

Multiphase materials such as colloids, emulsions, polymer or
surfactant solutions, wet granular systems and suspensions of deformable
objects like red blood cells  are characterized by a complex mechanical
behaviour  \cite{Larson1999}, due to the interaction of their
constitutive entities. The concentration is one of the key
parameters which control the rheology, determining especially the
transition from liquid-like to solid-like properties
\cite{Saint-Jalmes1999}.

Amongst these complex fluids,
liquid foams provide a
convenient model experimental system for laboratory studies of the
interplay between
structure, concentration and rheology.
This is because the bubbles which constitute
the foam's internal structure can be easily visualised
and manipulated.
The mechanical behaviour of foams
is very diverse: they appear
elastic, plastic or viscous depending on the deformation and
velocity gradient
\cite{Weaire1999,cohen-addad2005}.

A liquid foam consists of gas bubbles separated by a connected network of liquid
boundaries. This liquid phase occupies a
fraction $\Phi$ of the volume of the foam. The ``dry foam" limit,
in which $\Phi$ tends to zero, corresponds to polyhedral
bubbles separated by thin walls. It
is associated with a divergence of certain contributions to the viscous
dissipation \cite{buzza}. However, the foam's non-dissipative
properties (such as  surface energy
\cite{graner_2001}, shear modulus
or yield stress
\cite{Princen1983,Khan1986})
usually    tend to a regular,
finite limit when the liquid fraction $\Phi$ tends to zero.

The { total} ``yield drag" $F_Y^t$ is
the minimal force observed when there exists
(or, equivalently, required to
create)   a movement of the foam relative to an obstacle \cite{dolleteqrag05}.
It is  a
global,  geometry-dependent quantity directly measurable
in experiments and in practical applications of foams,
for instance
when a foam flows through a porous medium
\cite{porous}, or when one introduces an object into a foam (analogous
to sticking one's finger into shaving cream).

In the low-velocity limit (in which viscous dissipation is 
neglected \cite{cantatd05,denkov})
the total yield drag  $F_Y^t$
has two contributions. These are due to the  pressure
inside the
  bubbles, denoted $F_Y^p$, and the network of bubble
walls (i.e., soap
films with surface tension), $F_Y^n$.
Thus
\begin{equation}
F_Y^t = F_Y^p + F_Y^n.
\label{sum}
\end{equation}
Here we consider the network contribution $F_Y^n$ and
show how it is affected by the liquid content of the foam.

We consider a single layer of { equal-area} bubbles to facilitate
preparation and analysis of experiments, as well as  numerical and
analytical  modelling
  \cite{FRIT}.
Section \ref{Sec:exp_methods} presents a 2D flow of a quasi-2D foam (a bubble monolayer)
  around a fixed circular obstacle
within a long channel: this is the historical experiment of
Stokes, already adapted to foams both in 2D
\cite{dolleteqrag05,asipauskas} and 3D 
\cite{alonso,bruyn04,pitois,pitois2} flows. We compare them with  truly 2D
simulations using two physically equivalent but differently optimised 
software packages (Section
\ref{Sec:simulations}). The simulation methods allow easy
variation of the geometrical parameters such as bubble, obstacle
and channel size and better control of bubble area. In Section
\ref{Sec:fluid_fr}, we discuss the issue of a common,
unambiguous definition of liquid fraction for all systems, in
theory, experiments and simulations. Section \ref{results}
presents our results: we show that the yield drag displays the
expected dependence
   with the bubble,
obstacle and channel size, and increases  when the liquid fraction
$\Phi$ decreases.  The  discussion  in Section \ref{discussion}
emphasises that taking into account the effect of liquid fraction
allows all data to be plotted on a single master-curve and that, although they cover different ranges of $\Phi$, the results of both simulation and experiment are consistent with a simple model.

\section{Experimental Methods} \label{Sec:exp_methods}

\subsection{Foam channel}
\label{foam_channel}

Our  bulk soap solution is de-ionised
water with 1\% Teepol, a commercial dish-washing liquid. Its surface
tension, measured with the oscillating bubble method (that is, imaging the interface shape), is
$\gamma = 26.1\pm 0.2$ mN m$^{-1}$, and its kinematic viscosity, measured with
a capillary viscometer, is $1.06\pm 0.04$ mm$^2$ s$^{-1}$.

\begin{figure}
\begin{center}
\centerline{
\includegraphics[height = 0.8cm,angle=0]{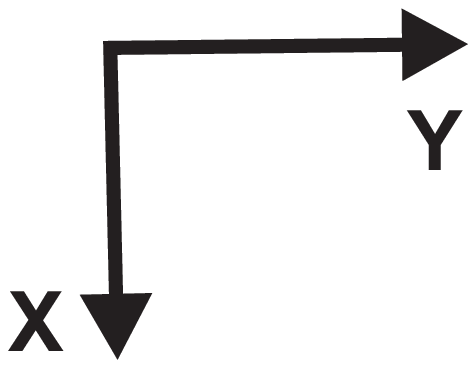}}
\centerline{
\includegraphics[height = 5cm,angle=-90]{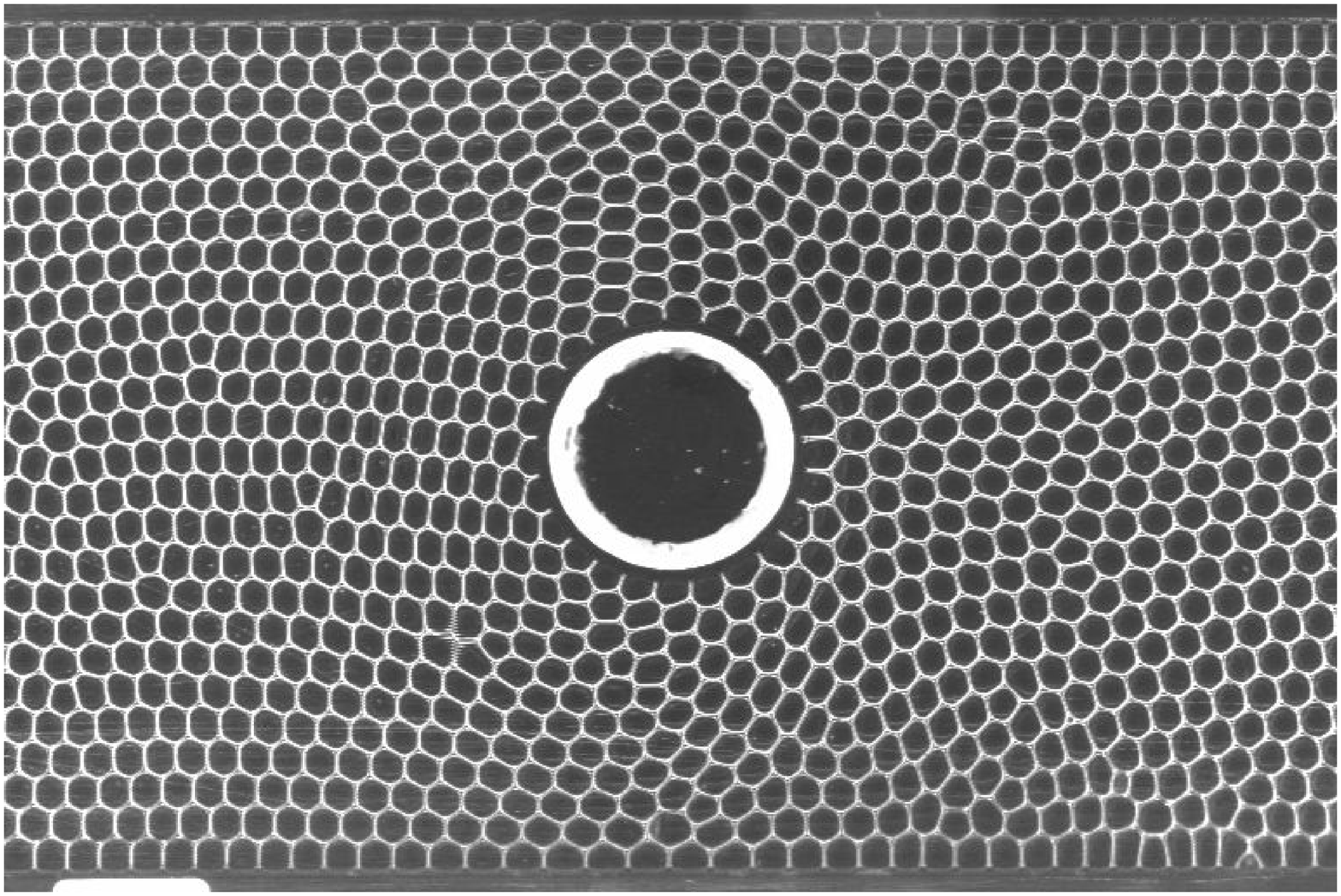}}
\end{center}
\caption{
Image of the experiment, with the foam confined between liquid and glass and flowing from top to bottom. Foam thickness $h=4.5$  mm; bubble area $A=16$ mm$^2$; obstacle diameter $d_0=3$ cm;
mean velocity $v=5.6$  mm s$^{-1}$; effective liquid fraction $\Phi =
0.06$.}
\label{image_exp}
\end{figure}

The experimental set-up \cite{dolleteqrag05} confines the foam
between a liquid reservoir and a glass lid (``liquid-glass" set-up \cite{vazc05}).
 A 1 m long,  $w_c=10$ cm wide tank
is filled with soap solution, leaving below the glass lid a
free space of thickness $h$ which we can adjust. We will call this
parameter the ``foam thickness" for simplicity. At its centre is a
circular obstacle of diameter $d_0=3$ (Fig.
\ref{image_exp})  or 4.8 cm. At the entrance to the channel, nitrogen is
blown at a computer-controlled flow rate, which varies between 5
and 500 ml min$^{-1}$. A typical value of the average velocity is
3 mm s$^{-1}$, for a 3.5 mm thickness and a
     flow rate of 50 ml min$^{-1}$.

The resulting foam consists of a horizontal monolayer of bubbles.
It  exits freely at
atmospheric pressure at the open end of the channel, $P=P_{atm}$.
In the absence of the obstacle, it yields a  two-dimensional  plug flow.
With the obstacle present, the flow
remains two-dimensional (even though the foam itself is
not exactly 2D \cite{vazc05}): there is no vertical component of the velocity.

Due to the presence of the obstacle, there is a velocity gradient.
There are many bubble rearrangements (so called ``T1s"
or neighbour-swapping events):
two three-fold vertices
contact, merge and re-separate.
We observe well-separated T1s; that is, between two T1s, there is  enough time
for the foam to relax to an equilibrium state.
The present  flow is slow enough that we can extrapolate the results to the 
low velocity limit
\cite{dolleteqrag05}, where a comparison with quasi-static calculations and simulations makes sense
\cite{weaire_qstat}.
(Although note that this is distinct from the zero-velocity case, that is the absence of flow.) The data presented below are all in this limit, extracted from the experiments as described in Appendix \ref{Sec:force_meas_exp_1}.

The bubble walls  meet the solid boundaries of the foam
(glass plate, lateral
channel walls, obstacle itself)  at a 90$^\circ$ angle \cite{footnote1}.
The surface density of bubbles is $1/A$, where
     $A$ is the average area per bubble (including its walls).
The foam is monodisperse: the area variation at the
channel entrance is less than 5\%.
The average area is fixed at a value ranging from
0.121 ${\rm cm}^2$
      to 0.393 ${\rm cm}^2$; most experiments have $A = 0.160$ cm$^2$.
Despite the low velocity, and hence the long transit time, we detect neither bubble coalescence nor coarsening.
  The effect of foam
ageing on rheology \cite{Gopal2003,Cohen-Addad2004} is thus
negligible.

   \subsection{Force measurements}
\label{forceexp}

\subsubsection{Total yield drag}

The obstacle
floats just below the top glass surface and is free to move,
without solid friction. However, it is linked to a fixed base
through a calibrated elastic fibre.
We track the obstacle displacement from its position at
rest using a CCD camera which images the foam flow from above.
We thus measure the force
exerted by the flowing foam on the obstacle (precision better than
0.1 mN)  \cite{dolleteqrag05}.

We check that the lift (spanwise component of the resultant force) is
consistently zero, within fluctuations, as
expected by symmetry  (data not shown). After a transient, the total
drag  $F^t$
(streamwise component of the resultant force) fluctuates around a
steady value:
we record the average and standard deviation of these steady flow data.
The extrapolation to the low velocity limit
(or zero-velocity intercept) of the force-velocity curve defines
the yield drag $F_Y^t$. It is independent of the bulk solution
viscosity \cite{dolleteqhag05}, and increases with the obstacle to
bubble size ratio \cite{dolleteqrag05}.

In this paper, we reanalyse the data already published in
\cite{dolleteqrag05} at various bubble areas, and we present new
data for another control parameter: the foam thickness. These data
are presented in Appendix \ref{Sec:force_meas_exp_1}. As explained
in Sec. \ref{phiexp}, the foam thickness provides a means by which to vary the
liquid fraction in experiments.

   \subsubsection{Network contribution to the  yield drag}
\label{expnetwork}

We measure $ F_Y^n$ as follows. Each bubble wall in
contact with the obstacle pulls it with a force equal to its line
tension $\lambda$ (the energy per unit length, which is of order $2 \gamma h$, see Appendix \ref{Sec:force_meas_exp_2}).
The elastic contribution  of the wall network to the drag
is then the vectorial sum of all these individual forces, which all have the
same modulus  $\lambda$.
As mentioned above, in a
quasi-static  flow each wall touches the obstacle at 90$^\circ$ angle.
    Thus it
suffices to find the contact points between bubble walls and the obstacle
and sum all outward normal vectors to the obstacle vectorially at
these contact points
(which is easy to determine for a circular obstacle).
If the downstream geometry of the foam was the same as that upstream, the
drag would be zero. Since bubbles are squashed upstream and stretched
downstream, the asymmetry means that there are more bubble walls pulling the obstacle
downstream,
and we measure a downstream elastic
contribution to the drag, $ F_Y^n/\lambda$.

The actual value of the line tension $\lambda$ is unimportant in what follows,
where only measurements of   $ F_Y^n/\lambda$ are compared.
However,  as presented in detail in Appendix \ref{Sec:force_meas_exp_2},
we measure $\lambda$ to check the
consistency of the orders of magnitudes of the independent measurements
of $ F_Y^n$ and $ F_Y^t$.

\begin{figure}
\centerline{{\bf(a)} }
\centerline{
\includegraphics[width=4cm, angle=0]{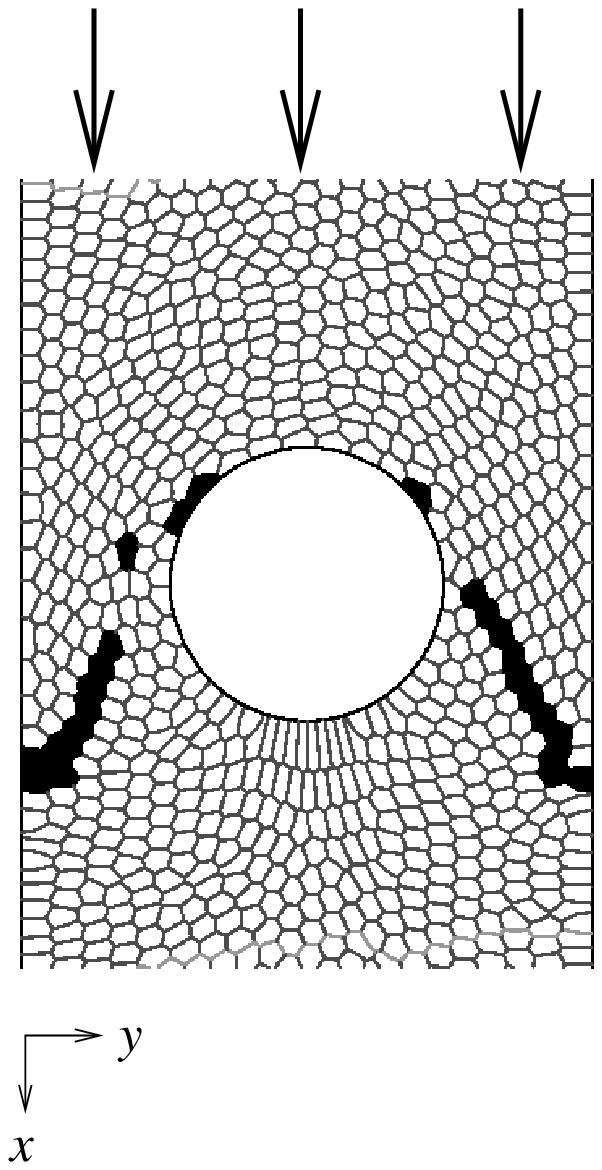}
}
\centerline{{\bf(b)} }
\centerline{
\includegraphics[width=3.5cm, angle=180]{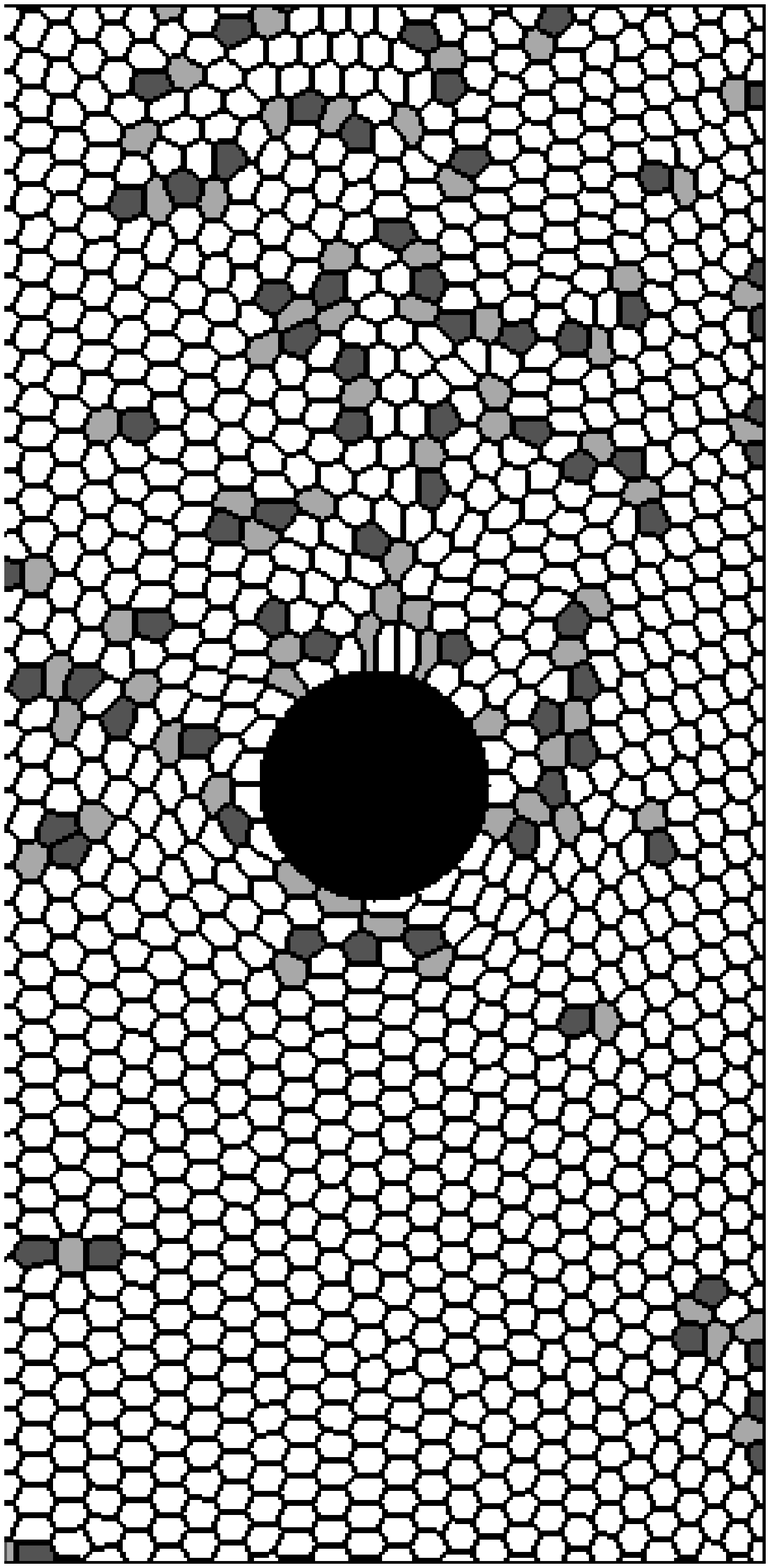}
}
\caption{
Images of simulated foam flow.
The $x$-axis is parallel to the flow
along the channel, with periodic boundary conditions (exiting
bubbles re-enter); axis $y$ is spanwise, with free-slip rigid
boundary conditions on either side of the channel.
(a) Surface Evolver.   The image shows the whole simulation domain of 750 bubbles; the shaded bubbles started in a horizontal
line. Here $d_0 = 4.8$ cm, $A = 0.16$  cm$^2$, $w_c = 10$ cm, ${L}_c=0.05$ cm and therefore $\Phi=0.0037$.
(b) Potts model.
The image shows   the simulated channel's  full width (except for a
few pixels)  of 256 pix, and half its length. Here
$d_0 = 74$ pix, $A = 100$ pix$^2$ and $\Phi=0.005$.
Bubbles coloured in white
    are without topological defect:
6-sided bulk bubbles, or 5-sided bubbles touching a lateral wall or
the obstacle
\cite{graner_2001}. Bubbles with fewer neighbours are in dark grey,
bubbles with more are in light grey.}
\label{Image_simul}
\end{figure}

\section{Simulations} \label{Sec:simulations}

\subsection{Deterministic simulations (the Surface Evolver)}

The Surface Evolver \cite{brakke92,simon_AERC}
offers the possibility to reach a true  quasi-static limit, that is a
succession of exact equilibrium states, through a deterministic
 minimisation of the foam's energy. It yields precise details of the foam
structure.

\subsubsection{Preparation of the foam}
    \label{prep_evolv}

  We use a mode in which  all bubble walls are
represented as circular arcs. The Surface Evolver lets these circular
arcs evolve in order to minimise the total perimeter (equivalent to
the energy, up to the prefactor $\lambda$). It enforces
the constraint that bubble areas $A$ remain fixed and determines the
corresponding Lagrange multipliers, namely each bubble's pressure $P$.
Since we can freely choose the units, we call them ``cm" and
use bubble size $A = 0.16$  or   0.353
    cm$^2$, channel width $w_c= 10$ cm, and obstacle diameters $d_0= 1.5$,
    3  and 4.8 cm,
to reproduce actual experiments.

The lateral sides of the channel  are  rigid and do not interact with the foam,
ensuring free-slip boundary conditions for the flow, { resulting in
a $90^\circ$ angle where a bubble wall meets the side}.
We adopt a periodic boundary condition
in the direction of motion:  bubbles that exit at the end of the
channel are fed back into the entrance of the channel.  We stop the simulation
when each bubble has passed the obstacle no more than once.

We begin with a rectangular lattice of $30\times 25$ monodisperse
bubbles of area slightly larger than the required area $A$. We
randomly perturb this
lattice  so that all the unstable four-fold
vertices dissociate into pairs of three-fold vertices and the whole foam
structure relaxes towards
equilibrium.
We then choose one bubble to be the circular obstacle, and slowly increase
its area to the required value (and correspondingly reduce  the
bubble areas to
$A$)
and constrain its edges to lie on a
circle.
The centre of the circular obstacle is then moved to the
centre of the channel and the structure again relaxed to
equilibrium.

\subsubsection{Simulation of the flow}

With the obstacle in the desired location and the foam close to equilibrium,
we start the quasi-static iteration procedure. This requires that we move the
foam past the obstacle, in a direction which we denote by $x$; the
difficulty is in
doing this with the periodic boundary conditions without fixing any vertices or
bubble shapes. Our method is to choose a continuous line of
consecutive bubble walls
from one side of the channel to the other. Joining this to a line at $x=0$
with lines along the channel walls defines a plane region with a certain area, which we constrain.
At each iteration  we choose a convenient line of consecutive  walls, and
increment the target area of
the region formed by a small amount $dA$ (equal to 0.05 cm$^2$
     in all simulations), resulting in a slight movement of a line of
films without modifications to
the bubble areas. The total perimeter of the structure is then reduced
until it converges to a constant value (Fig. \ref{Image_simul}a),
so that measurements can be performed.

We have double-precision values for
the network geometry.
We measure the  network contribution $F_Y^n$ to the yield drag as in
experiments
(section \ref{expnetwork}).
It is the sum of the
unit vectors of the bubble wall  with one end attached to the
obstacle, expressed in
units of the line tension (hence as a dimensionless number).
Here too, we
check that the lift is
consistently zero within fluctuations (data not shown).

\begin{figure}
\psfrag{F/lambda}{\hspace{-8mm}Network Force $F_Y^n/\lambda$}
\psfrag{Wall Network Force/lambda}{Network Force $F_Y^n/\lambda$}
\centerline{
\includegraphics[height = 7.5cm,angle=270]{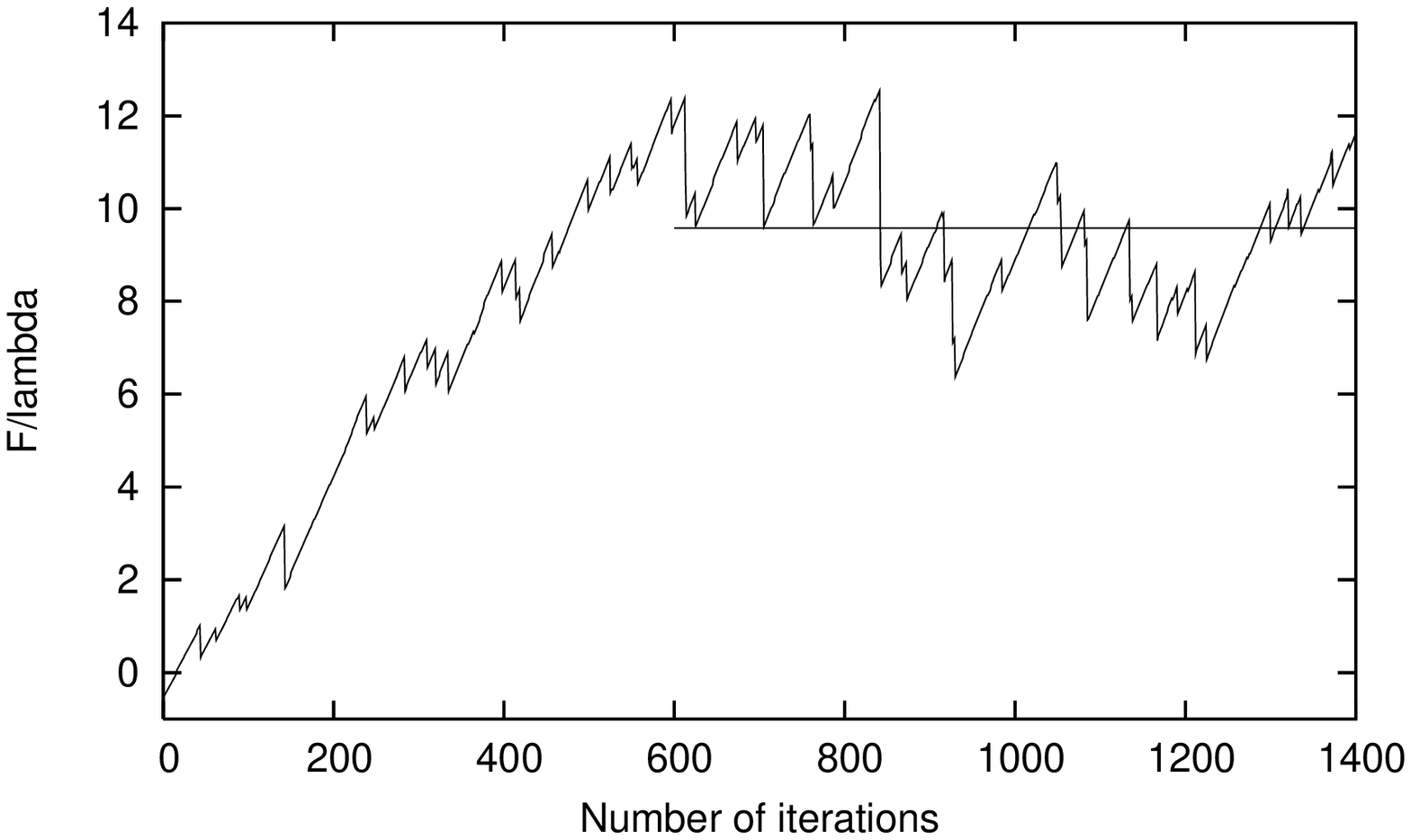}
{\bf(a)}
}
\centerline{
\includegraphics[height = 7.5cm,angle=270]{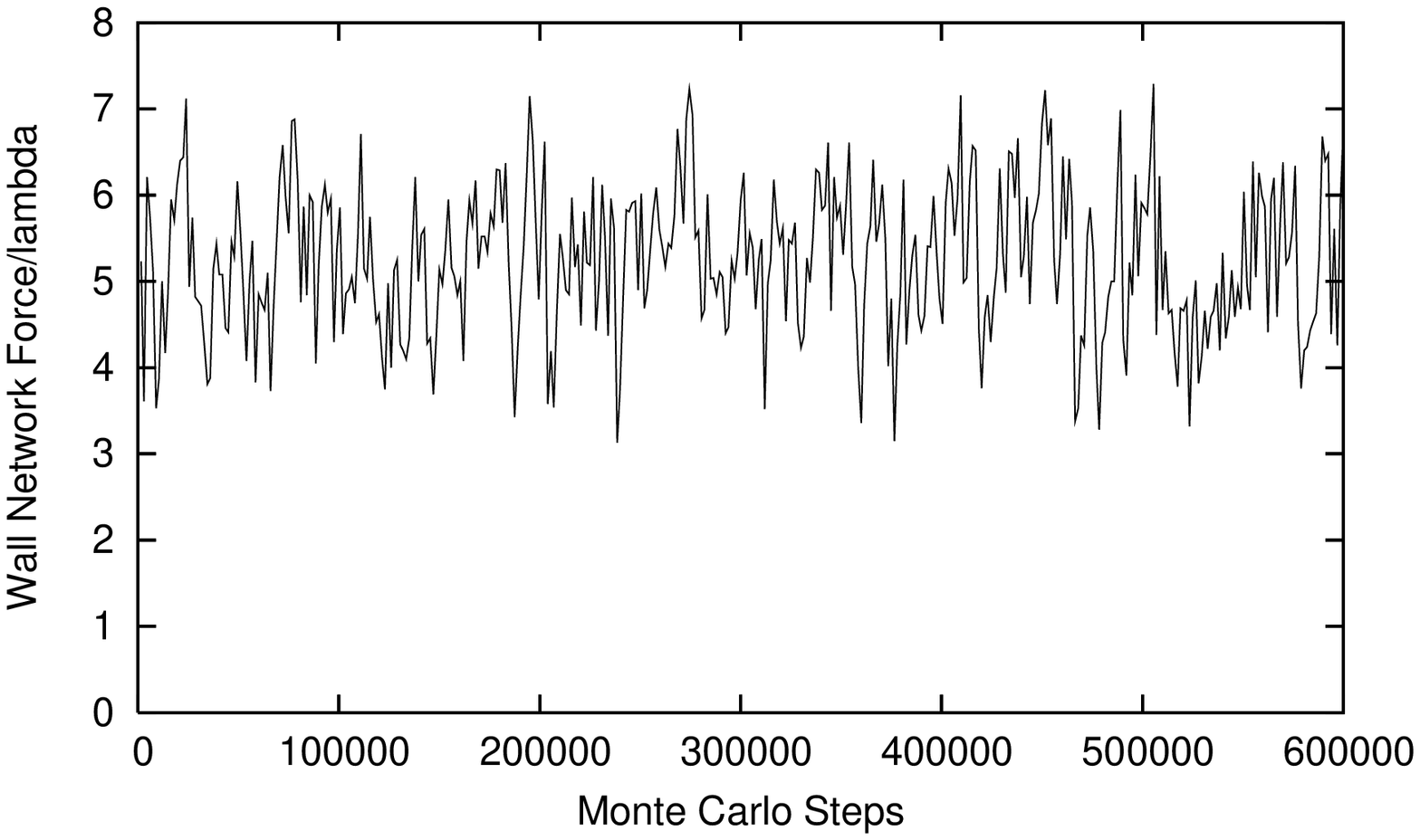}
{\bf(b)}
}
\caption{
The network force $F_Y^n$ (expressed in units of the line tension $\lambda$) measured in simulations {\it versus} time.
(a)
Surface Evolver data plotted every iteration step;
$d_0 = 4.8$ cm, $A = 0.16$ cm$^2$ and ${L}_c = 0.05$ cm, $\Phi=0.004$. The plateau
value is $F_Y^n = 9.6 \pm 1.4$.
(b)
Potts model data plotted every 1500 Monte Carlo Steps;
$d_0 = 74$ pix, $A = 100$ pix$^2$, $\Phi=0.005$. The plateau
value is $F_Y^n = 5.4 \pm 1.1$.
}
\label{time}
\end{figure}

With the area increment $dA=0.05$ cm$^2$, the transient lasts for about 600
iteration steps (Fig.
\ref{time}a). This is comparable to, but still smaller than, the total
simulation time that is reasonably accessible.
     After this
transient, the drag fluctuates around a steady value. Such fluctuations,
due to  the
rearrangements of the bubbles,
 recall the stress drops observed in
Couette experiments for disordered foams
\cite{Lauridsen2002,dennin}.
We   record the average and standard deviation
of these plateau (steady-flow) data for a total of
1500-600 = 900 iterations. To validate the choice of our simulation size, we
checked once that the drag forces are the same with more bubbles
in the direction of flow (1250 bubbles instead of
750), although the transient is longer.

Each simulation takes about   35 hours on a    Pentium IV  3.20
GHz processor: typically
a several
hour build-up to the initial structure (inflating the obstacle), plus
 one iteration per minute (depending on
the number of bubbles and on the liquid fraction).

\subsection{Stochastic simulations (Potts Model)}

To simulate a larger
number  of  bubbles, the Potts model adapted for
foam rheology \cite{jiangpre} also
minimises the same energy, but stochastically (Monte-Carlo),
which increases the simulation speed. It thus provides more
statistics on $F_Y^n$ and allows quicker variation of the geometrical parameters.

\subsubsection{Principle of the Potts Model}

The Potts model   is derived
from a large-$Q$ Potts model run at zero temperature,
a model widely  used to model grains in crystals \cite{srolovitz}.
It has been also applied to different domains of foam physics,
including rheology, by enforcing the conservation of bubble size and
applying an external force   \cite{jiangpre}.

We consider a 2D square lattice. Each site $i$ has an  integer index
$\sigma_i$.
The  $k^{th}$ bubble is defined as the  domain consisting of  all
sites with the
same  index value $\sigma_i = k$.
Thus bubbles tile the plane without gaps or overlaps. The evolution
is driven by the
minimisation of a total energy ${\cal H}$ (strictly speaking, it is a
Hamiltonian), which has the same three physical
ingredients
as in the Surface Evolver: 
interfacial energy, area constraints, external forcing of the flow.
Since the calculations are performed on a lattice, we have
\begin{eqnarray}
{\cal H} & = & \lambda \sum_{{i,j} \; \rm neighbours}  \left[ 1 -
\delta(\sigma_i,\sigma_j)\right]
\nonumber \\
&& + \chi \sum_{{\rm bubbles} \; k} \left( A_k  - A_k^t\right)^2
+ b \sum_{{\rm sites} \; i}  \; x_i
.
\label{hamiltopotts}
\end{eqnarray}

The first term represents the contribution of the energy
of the interfaces between the bubbles. Minimising this term leads to
perimeter minimisation. Here $\delta$ is the Kronecker symbol:
$1-\delta$ is equal to 1 if  the neighbouring sites $i$, $j$
belong to different bubbles  ($\sigma_i \neq \sigma_j$); else it
equals zero. 
We choose to evaluate this term with the fourth nearest neighbour
interactions to obtain an isotropic line tension insensitive to the details of the lattice
\cite{holm}. 

The second term
keeps each bubble
area $A_k$ (the number of sites with the same index) close to its
predefined target value $A_k^t$.  Here $\chi$ is the
compressibility, which we choose to be high enough to keep bubble areas
constant to within a few pixels.
The balance between this term and the preceding one simulates a foam
relaxing towards  mechanical equilibrium.

The third term is a bias term that describes an energy gradient,
hence a homogeneous
external force field.
Here $b$ is the  bias intensity and $x$ the site's coordinate along
the flow.    Without obstacle, the resulting velocity profile  would
be a plug flow.

We use a Metropolis algorithm to evolve the foam: we randomly select
a site at a bubble boundary,
change its index to the value of a neighbour if and only if this
decreases the total energy
(eq. \ref{hamiltopotts}).
Several independent changes  are tried successively;
a Monte Carlo Step (MCS)  is  defined conventionally as a number of
tries equal to
the total number of lattice sites.

\subsubsection{Simulation of the flow}

As for the Surface Evolver
(section \ref{prep_evolv}),
we choose
a periodic boundary condition in the direction of flow and free-slip rigid
boundary conditions on the channel sides.
To ensure that it
    does not affect the steady-state measurements presented below,
  the total channel length is $4w_c$, out of which  only $2w_c$ are
used for measurements and are shown on
Fig. (\ref{Image_simul}b).

To match the experiments, we choose
     $16 \le d_{0} \le 148$ pix, $64 \le A \le 400$    pix$^{2}$ and
$64 < w_c <  512$ pix.
Initially, we insert a rigid round obstacle in the centre of the
channel, and let
a perfectly ordered foam (honeycomb pattern) flow
in.  We
then switch off the bias term by setting   $b=0$, and relax the foam to
ensure that the
bubbles recover their (near) equilibrium state.
The foam has reached the stationary state at the end of this  preparation.

We then switch the bias on again, and use the smallest
bias $b$ for which the foam flows, which is constant
and independent
of parameters such as bubble diameter.
We perform measurements at intervals of 1500 MCS (during  which a bubble   moves a
few pixels).


We measure the  network contribution to the drag using the same
method as in the experiments and Surface Evolver simulations. It
fluctuates around a steady value:  we record the average and
standard deviation of these plateau (steady-flow) data (Fig.
\ref{time}b). We run each simulation for a total of 600,000 MCS,
during which  a bubble passes completely through the channel  but
no bubble passes the obstacle twice. One simulation takes about 12
hours  on a Pentium IV 2.8 GHz processor.

\section{Liquid fraction} \label{Sec:fluid_fr}

In ideal 2D foams (Sec.
\ref{phitheo}), given $A$, 
 the liquid fraction scales as the square of the vertex radius.  This radius in turn relates to a cut-off length $L_c$, that is the length  at which a bubble edge becomes unstable and the
surrounding bubbles undergo a T1.
This length $L_c$ is always defined, and is relevant to the mechanical
properties investigated here.
In experiments, $L_c$ depends on the actual (3D) shape of bubbles;
in simulations, $L_c$ is an input parameter.

It suggests a
  \emph{coherent} definition of the
\emph{effective} liquid fraction, presented below, consistent within and between experiments,
simulations and theory.
Note that  we consider here a foam dry enough to 
have a non-zero shear modulus \cite{Bolton1990},
that is, below the critical 
liquid fraction \cite{Princen1983} (see eq. \ref{Phi_c}
for the honeycomb value).

\subsection{Ideal 2D foams} \label{phitheo}

\begin{figure}
\begin{center}
\includegraphics[width=6cm]{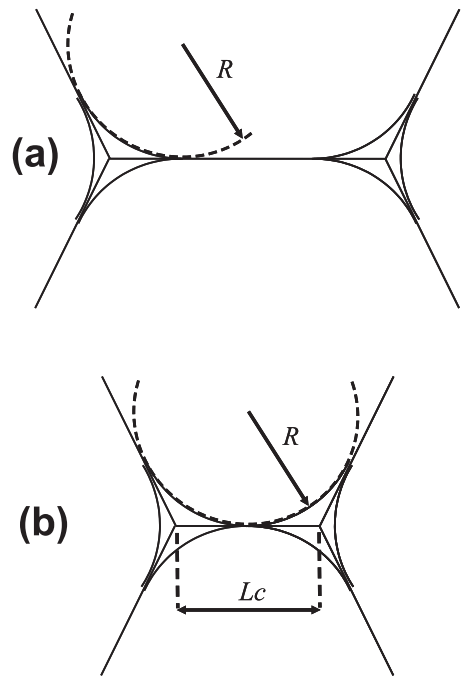}
\end{center}
\caption{Picture of two adjacent three-fold vertices with Plateau
borders. (a) We apply the decoration theorem \cite{Bolton1990}  to
model a  wet foam. The liquid is present only at the vertices, and
(assuming here straight or nearly-straight walls) the uniformity
of pressure $P$ inside bubbles forces each gas/liquid interface
to have the same radius of curvature, $R$. (b) Critical position
of the vertices
   just before the ``T1"  neighbour-swapping event. It
defines the cut-off wall length ${L}_{c}$. } \label{vertex}
\end{figure}

In an ideal 2D foam, a Plateau border  is a triangle with concave
edges of radius $R$ which match tangentially three straight lines
meeting at 120$^\circ$ (Fig. \ref{vertex}). The  area $A_{PB}$ of a Plateau
border is
\cite{Weaire1999}:
\begin{equation}
A_{PB}=\left(\sqrt{3} - \frac{\pi}{2}\right) R^{2}.
\label{area_PB}
\end{equation}

The liquid fraction $\Phi$ is defined as:
   \begin{equation}
 A_l = A\Phi, \label{Ag}
\end{equation}
where $A_l$ is the area occupied by the liquid.
Since each bubble has $n$ Plateau borders, each being shared between 3 bubbles, we have $A_l = n A_{BP}/3$. For a honeycomb
array of bubbles, $n=6$ and:
\begin{equation}
\Phi = \left (2\sqrt{3}-\pi\right)\; \frac{R^2}{A}. 
\label{Phi_R}
\end{equation}

Here we consider foams where bubbles have the same area (monodisperse foams),
but not necessarily the same number of sides $n$ (topological disorder). Since 
on average over the whole foam $\bar{n} \approx  6$
 \cite{Weaire1999}, and since Plateau borders have almost the same
size and radius of curvature, eq.
(\ref{Phi_R}) still holds here approximately.

A T1  is triggered when the distance between these vertices
becomes smaller than a   cut-off wall length ${L}_c$, which
increases with $R$, and thus with $\Phi$. To make this observation
more quantitative, one possible convention to define ${L}_c$ is
the condition that two vertices touch
   (Fig. \ref{vertex}):
\begin{equation}
\frac{R}{\sqrt{3}} =  \frac{{L}_c}{2}, \label{ellc_R}
\end{equation}
so that, together with  eq. (\ref{Phi_R}):
\begin{equation}
\Phi   = \frac{3}{2} \left( \sqrt{3} - \frac{\pi}{2}\right)
\frac{{L}_c^2}{A} \approx 0.242 \; \frac{{L}_c^2}{A}.
\label{eq:se_phi}
\end{equation}
Given $A$, the physical information conveyed by $R$, $\Phi$ or
${L}_c$ is the same. For
comparison between different experiments or simulations,
   we  use $\Phi $ because it is dimensionless.

For an ideal honeycomb without shear, all
vertices merge at the same liquid fraction: the hexagons become circular
when ${L}_c$ equals the side-length of the hexagons;
the bubbles are circular, with a radius equal to $R$. This
critical liquid fraction is \cite{Princen1983}:
\begin{equation}
\Phi_c =  1 - \frac{\pi}{2\sqrt{3}} = 0.0931. 
\label{Phi_c}
\end{equation}

\subsection{Experiments} \label{phiexp}

In the experiments, the actual (3D) shape of bubbles
is determined by the foam thickness.
 Fig. (\ref{image_exp}) shows a foam
thickness of 4.5 mm; beyond this thickness, the bubbles undergo a
three-dimensional instability and the foam is no longer a
monolayer \cite{coxwv01}. At the other extreme, at 2 mm and below, the
bubbles are circular and separated: both the foam's 2D shear
modulus and the yield drag vanish.
When $h$ increases, $L_c$ decreases, thus $\Phi$ decreases too
(Fig. \ref{Manip_force} in Appendix \ref{Sec:force_meas_exp}).

We show in Appendix
\ref{Sec:Appendix_model} that there is a correspondence between
$L_c$ and the length $L_{\max}$ of the
edge of a bubble just attached to the obstacle (Fig.
\ref{bubbles}c in Appendix \ref{Sec:Appendix_model}). Since $L_{\max}$ is much
bigger than $L_c$, it can be estimated with much more precision in
experiment (the uncertainty in both $L_c$ and $L_{\max}$ is one pixel).

We measure on
the skeletonized image the length $L_{\max}$  several times preceding its disappearance during a T1
process, and keep the average as $L_{\max}$ and the standard
deviation as $\delta L_{\max}^{\mathrm{disp}}$. The
skeletonization itself induces a systematic error in determining
the actual position of the vertex centre; we estimate as $\delta
L_{\max}^{\mathrm{skel}} = 1$ pixel the systematic error on
$L_{\max}$. The total uncertainty on $L_{\max}$ is therefore
$\delta L_{\max} = \sqrt{(\delta L_{\max}^{\mathrm{disp}})^2 +
(\delta L_{\max}^{\mathrm{skel}})^2}$.

To deduce $\Phi$ from the measurements of $L_{\max}$,
we combine Eqs. (\ref{Phi_R}) and (\ref{R(A)}) to get the
following expression as a function of $L_{\max}^2/A$ only:
\begin{equation}\label{Eq:Phi_exp}
\Phi = \frac{3}{2} \;  \frac{2\sqrt{3}-\pi}{2+\sqrt{3}} \; \left(
\frac{\sqrt{A}}{L_{\max}} - \frac{L_{\max}}{4\sqrt{3A}} \right)^2
.
\end{equation}
Its uncertainty is:
$$ \frac{\delta\Phi}{\Phi} = 2 \; \frac{\delta L_{\max}}{L_{\max}}
\frac{\sqrt{A}/L_{\max} + L_{\max}/4\sqrt{3A}}{\sqrt{A}/L_{\max} -
L_{\max}/4\sqrt{3A}} .
$$

\subsection{Simulations}
 \label{phievolver}

The Surface Evolver requires that we specify explicitly the cut-off wall length
${L}_c$ at which two three-fold vertices are allowed to contact,
merge and re-separate. 
 Since
$L_c$ is an input parameter, it is determined without uncertainty.
This defines explicitly an effective liquid fraction
 (eq. \ref{eq:se_phi}), at least for small values  of $\Phi$.
 We choose $ {L}_c $ to be of the
order of 0.1 cm or slightly smaller, reaching $\Phi=0.0015$, 0.0037, 0.0061 and 0.015.
 At very small  values of $\Phi < 6  \;  10^{-6}$, films behind the obstacle
would get very stretched and lead to numerical problems.
Attempting larger values of $\Phi >  0.015$ would lead to poor
convergence in the Surface Evolver and would require that we
simulate the actual geometry of the liquid in the vertices
(including 4-fold vertices).

\label{phipotts}

In the Potts model, 
 the cut-off distance ${L}_c$ at which two vertices merge
is either 1 or 2 pixels. We use this range to define the uncertainty on 
the value of $\Phi$. Since  we will plot the results in log scale,
we choose $L_c \approx \sqrt{2}$ and $\Phi   \approx  0.242 \times 2 / {A} \approx 0.5 {A}^{-1}$
 to lie in the middle of this interval.
 Thus 
the area $A$ of bubbles (that is, the number of pixels per bubble) defines an effective
liquid fraction, at least for small values  of $\Phi$.
The simulated range $64 $
pix$^{2} \le A \le 400$ pix$^{2}$ corresponds to  $0.00125  < \Phi
<0.0075 $, large enough  to describe 
realistically the shape of bubbles, and small enough to 
keep the computation time reasonable.

\section{Results}
\label{results}


{ The} experiments and both simulations present { qualitatively}
similar images
   (Figs.  \ref{image_exp},   \ref{Image_simul}) and consistent
results for the yield drag force, always directed downstream.

There are {\it a priori} four lengths in this problem: the channel
width $w_c$, the obstacle diameter $d_0$, the  bubble size
$\sqrt{A}$; and  the cut-off length ${L}_c$. As
far as we can tell, it is safe to assume that the channel length
(if long enough) is irrelevant here. These four lengths can be
reduced to three dimensionless parameters. We  present the
results using: $d_0/w_c$ which characterises the flow geometry;
$d_0/\sqrt{A}$ which describes the foam-obstacle interaction; and
 ${L}_c^2/A$ which characterises the threshold for T1
rearrangements, and corresponds to the liquid fraction $\Phi$.

\subsection{Effect of obstacle to channel size ratio}

\begin{figure}
\begin{center}
\centerline{
\includegraphics[height = 6cm,angle=0]{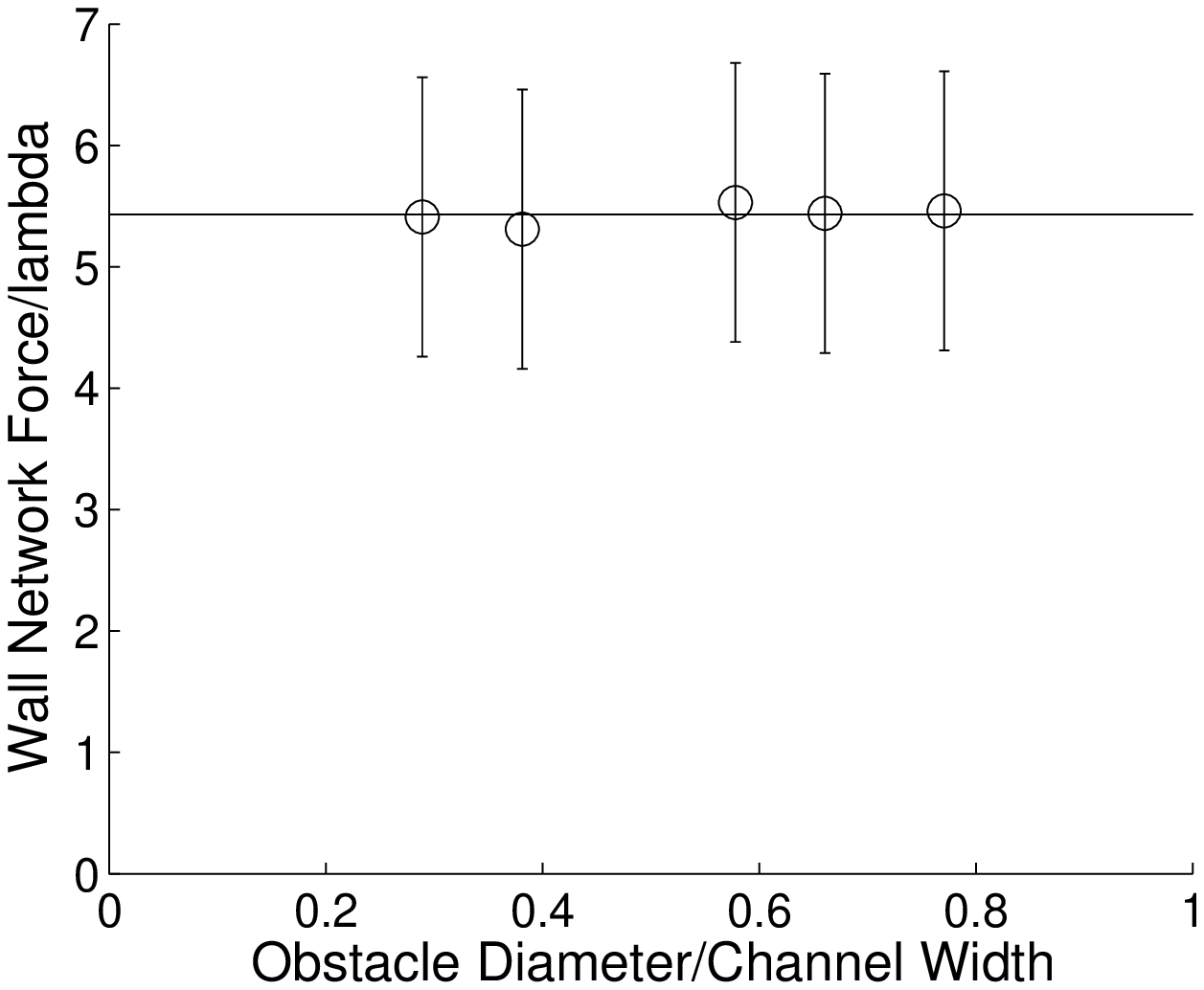}
}
{\bf(a)}
\centerline{
\includegraphics[height = 6cm,angle=0]{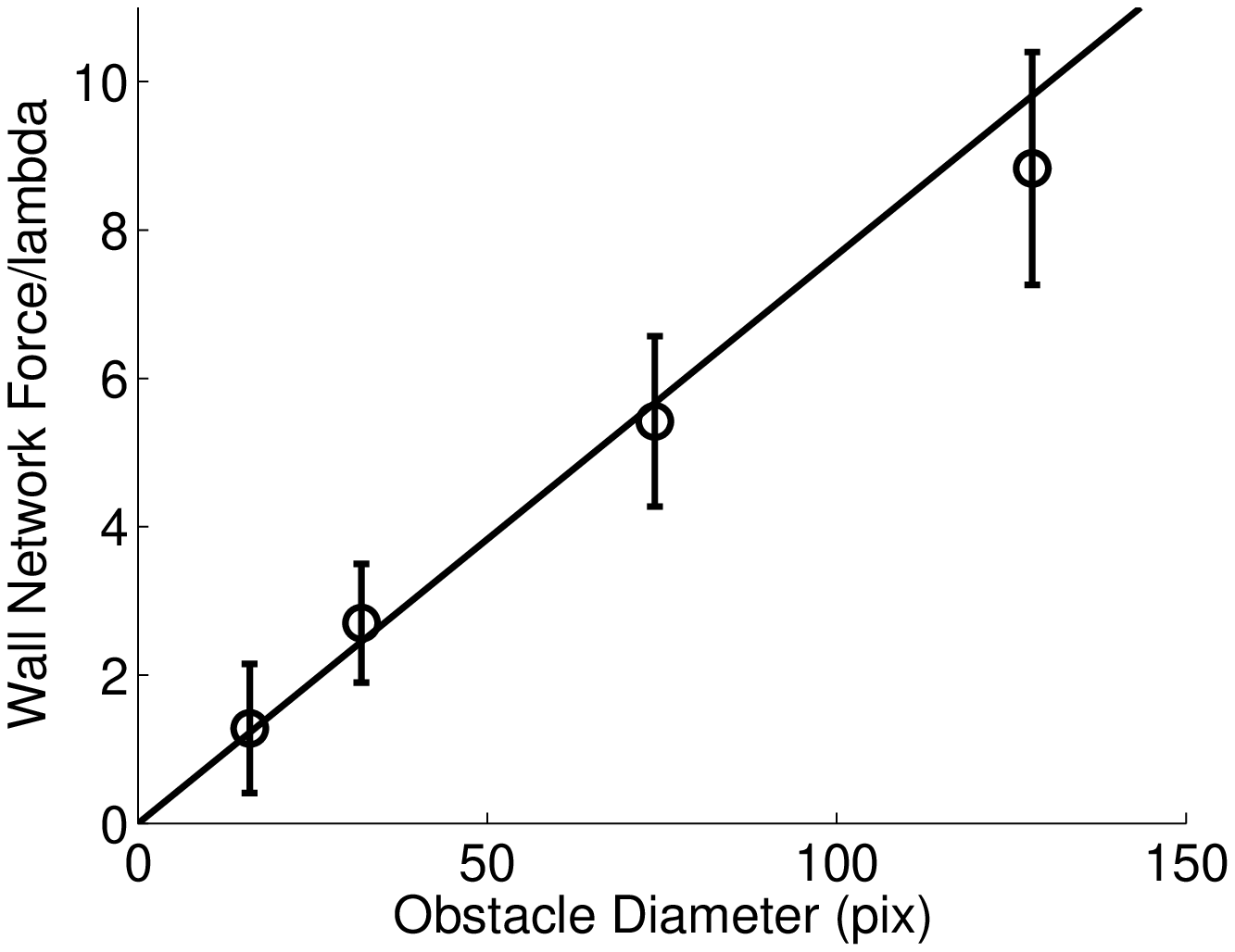}
}
{\bf(b)}
\end{center}
\caption{
Network contribution to the yield drag $F_Y^n$ (expressed in unit of
$\lambda$), measured in Potts
model simulations with $A = 100 $ pix$^{2}$ ($\Phi =0.005$).
(a) $F_Y^n$  {\it versus} $w_c$, for  $d_0=74$ pix;
the solid line is the average (value 5.43).
(b) $F_Y^n$ {\it versus} $d_0$, for $w_c=256$ pix;
the solid line is a linear fit with zero intercept,
$F_Y^n=0.77\; d_0/\sqrt{A}$.
}
\label{obst_channel}
\end{figure}

Potts model simulations indicate that the network yield drag is
independent of the ratio of obstacle size to channel width, $d_0/w_c$ (Fig. \ref{obst_channel}a).
This ceases to be valid at small $d_0$, 
when  the obstacle is comparable in size to
a bubble, and at large $d_0$, when the
distance between the obstacle and the channel side is small \cite{simon_AERC}.

The lack of dependence on $d_0/w_c$ that we find
characterises the yielding behaviour of the foam: it means that
only a small region near the obstacle is affected by the flow
\cite{bruyn04}. Nonetheless,
    the zone where the obstacle influences the flow is
larger in 2D \cite{dollet_local_en_preparation}  than in  3D
\cite{bruyn04}, as elastic or hydrodynamic interactions would suggest.

\subsection{Effect of obstacle to bubble size ratio}

Potts model simulations indicate that the network yield drag
    increases linearly with the obstacle size $d_0$ (Fig.
\ref{obst_channel}b) at fixed bubble area.
This is consistent with the force increasing as $d_0/\sqrt{A}$, also
suggested by the available Surface Evolver data,
as well as by  experimental measurements of the total force
that show the role of the
obstacle's spanwise dimension (``leading edge") \cite{dolleteqrag05}.
Note that most elastic properties
of a foam scale like $1/\sqrt{A}$ \cite{Weaire1999}. In fact,  when
$A$ increases, the density of bubbles and of bubble walls
decreases, and so does a foam's elastic modulus  (it would eventually vanish
if there were only one large bubble left).

\subsection{Effect of liquid fraction}

\begin{figure}
\psfrag{Effective liquid fraction xx}{Effective liquid fraction}
\psfrag{lc2/A}{$L_c^2/A$}
\psfrag{Force xx}{$F_Y^n \sqrt{A} / \lambda d_0$}
\begin{center}
\includegraphics[width=8cm]{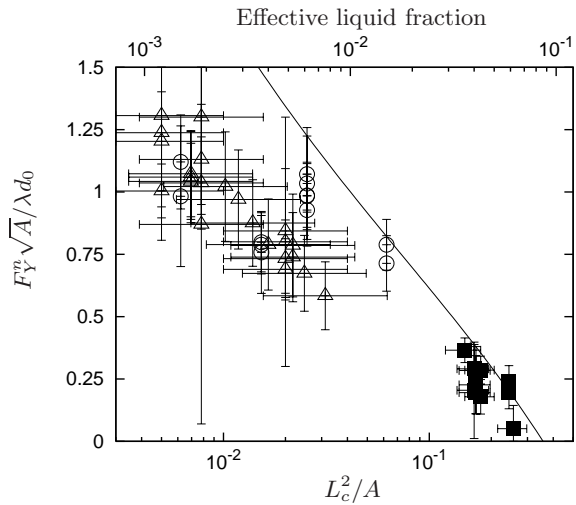}
\end{center}
\caption{The network contribution to the yield drag $F_Y^n$,
rescaled by  $\lambda d_0/\sqrt{A}$, is plotted {\it versus} the
dimensionless quantity $L_c^2/A$ (bottom scale), 
that is {\it versus}
the effective liquid fraction $\Phi $ (top scale), in the range $10^{-3} <
\Phi < 10^{-1}$. All control parameters ($d_0$, $w_c$, $A$ and $\Phi$) are varied.
 Vertical bars indicate the standard deviation of
the force fluctuations in time around the plateau value.
Horizontal bars indicate the uncertainty discussed in Sec.
\ref{Sec:fluid_fr}.  Data are from experiments
($\blacksquare$), Surface Evolver ($\circ$) and Potts model
($\vartriangle$); the solid line denotes the analytical model,
from eq. (\protect\ref{F(phi)}), without adjustable parameters.  Note that 
the horizontal scales are logarithmic and shifted with respect to each other.}
\label{general}
\end{figure}

We need to separate the effects of   foam geometry,
$d_0/\sqrt{A}$,  from those of liquid fraction, $\Phi$. We thus rescale
the network contribution to the yield drag $F_Y^n$ by
$d_0/\sqrt{A}$, and plot all our data as a function of $\Phi$. 
All the data, from both experiments and simulations, are well
rescaled   in the range { $10^{-3}
< \Phi <  10^{-1}$} (Fig. \ref{general}). This is the main result of the
present paper.

\section{Discussion}
\label{discussion}

\subsection{Model}

The effect of the liquid fraction on the network drag can be understood as
follows.
A bubble of area
$A$ detaches from the obstacle when its width is of order ${L}_c$, and thus
its length is of order $A/{L}_c$.
When $\Phi $ decreases, ${L}_c$
decreases too.
   Bubbles stretch more downstream, and more bubbles
   pack behind the obstacle.
The number of bubble walls pulling the obstacle downstream
increases; simultaneously, the number of walls upstream decreases.
This larger up/downstream asymmetry   results in an increase in
the resulting drag $F_Y^n$. The contribution from the network (or
bubble walls) increases as their number per unit length along the
obstacle boundary, namely ${L}_c^{-1}$, and thus scales like
$1/\sqrt{\Phi  }$.

However, the length of the region on which stretched bubbles act
decreases, and the divergence in $1/\sqrt{\Phi  }$ is in fact
softened by a geometrical factor. As shown in Appendix
\ref{Sec:Appendix_model}, we can estimate this factor by
integrating the bubble wall contribution around the obstacle.
When $\Phi$ increases, $F_Y^n$ decreases; it vanishes for
$\Phi=0.086$. This is close to the rigidity loss value (eq. \ref{Phi_c}).
Eq.  (\ref{F(phi)}) is plotted in Fig. (\ref{general}),
without adjustable parameters.
It shows qualitative agreement with the data over two decades of
liquid fraction, suggesting
that it  captures the essence of the physics.

\subsection{Influence of the control parameters}

In the limit of
low $\Phi $, the development of the above argument indicates that,
provided that the obstacle diameter and the obstacle-wall distances are larger than the bubble diameter,
$F_Y^n$ increases according to:
\begin{equation}
F_Y^n = \frac{0.516}{\Phi^{1/4}} \; \frac{\lambda \; d_0}{\sqrt{A}}.
\label{scaling_eq}
\end{equation}


In simulations, 
if we multiply
the bubble and obstacle diameters,
expressed in units of the
cut-off length,
by the same prefactor, the network drag changes (data not shown), due to the change in $\Phi$.

\begin{figure}
\psfrag{Wall Network Force/lambda}{Network Force $F_Y^n/\lambda$}
\begin{center}
\includegraphics[width = 8cm,angle=0]{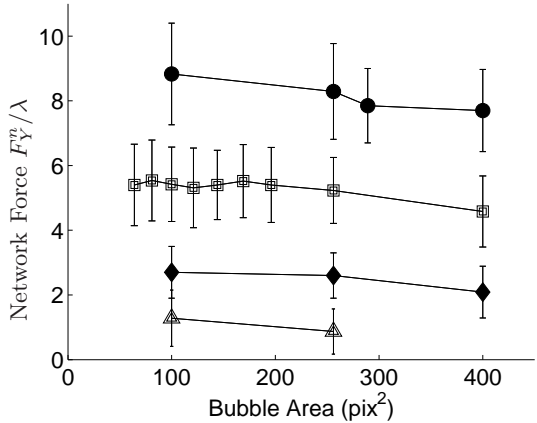}
\end{center}
\caption{
Opposite effects in  simulations  of $F_Y^n$  (here with Potts model).
When  $A /{L}_c^2$ increases,  both $d_0/\sqrt{A}$ and $\Phi\approx A^{-1} $ decrease;
so that  
$F_Y^n $ barely varies (eq. \ref{scaling_eq}). Obstacle diameter $d_{0}$
equal to 16 ($\vartriangle$), 32 ($\blacklozenge$), 74 ($\square$)
and 128 ($\bullet$).
}
\label{antagonist}
\end{figure}

Conversely, increasing only the bubble area $A$ at fixed ${L}_c$
    simultaneously decreases  both $d_0/\sqrt{A}$ and $\Phi  $.
This has two opposing effects, the former decreasing   $F_Y^n$,
the latter increasing it, resulting in an almost constant  $F_Y^n$ (Fig. \ref{antagonist}). (In fact there is a weak dependence on area, varying as $A^{-1/4}$.) 
This shows that the relevant way to vary
the liquid fraction in simulations is to modify $L_c^2/A$ at given
$d_0/\sqrt{A}$. \label{twoeffects}

\subsection{Saturation at low $\Phi$ }

\label{disc_limit_model}

\begin{figure}

\centerline{
\includegraphics[width=6cm, angle=-90]{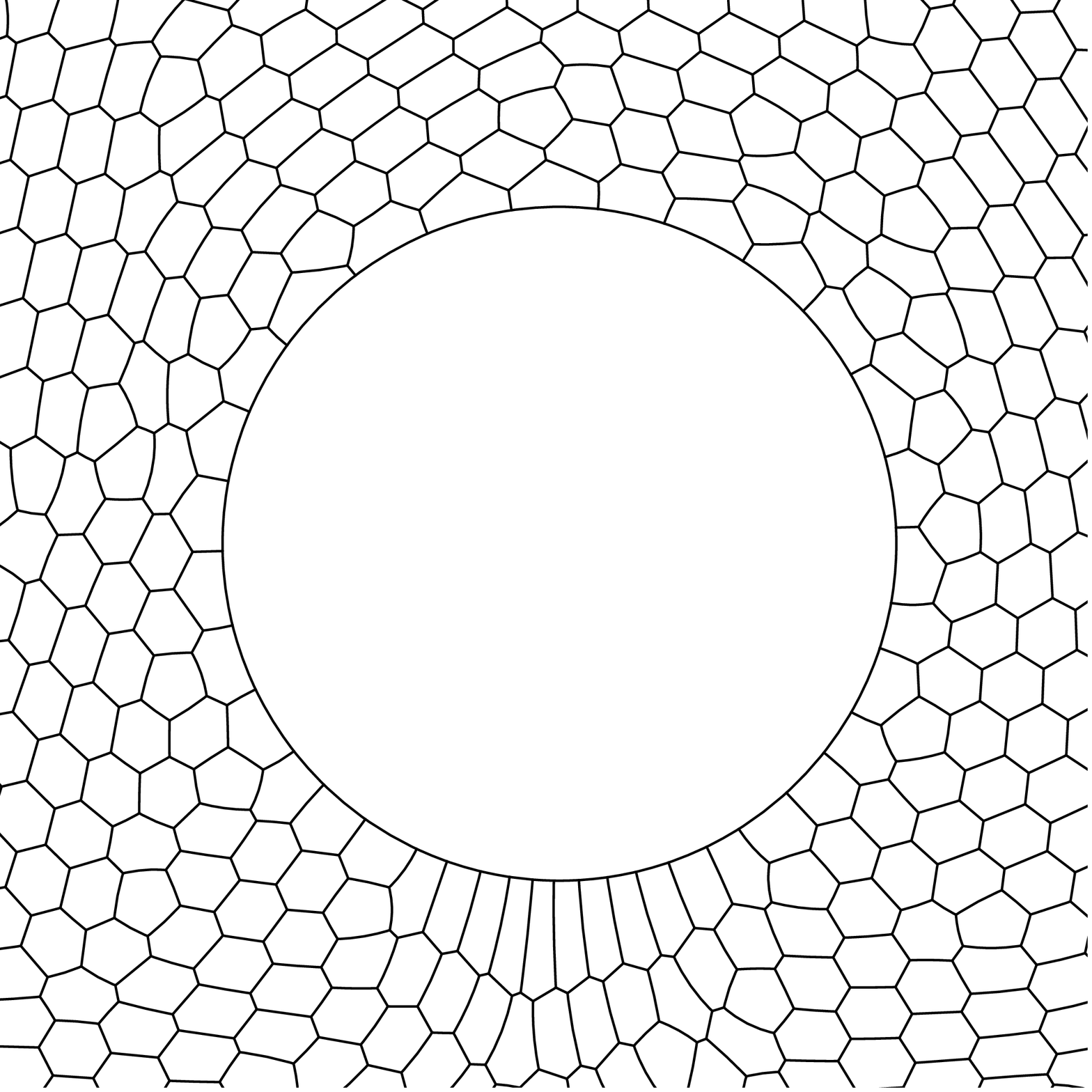}
{\bf(a)} }
\centerline{
\includegraphics[width=6cm, angle=-90]{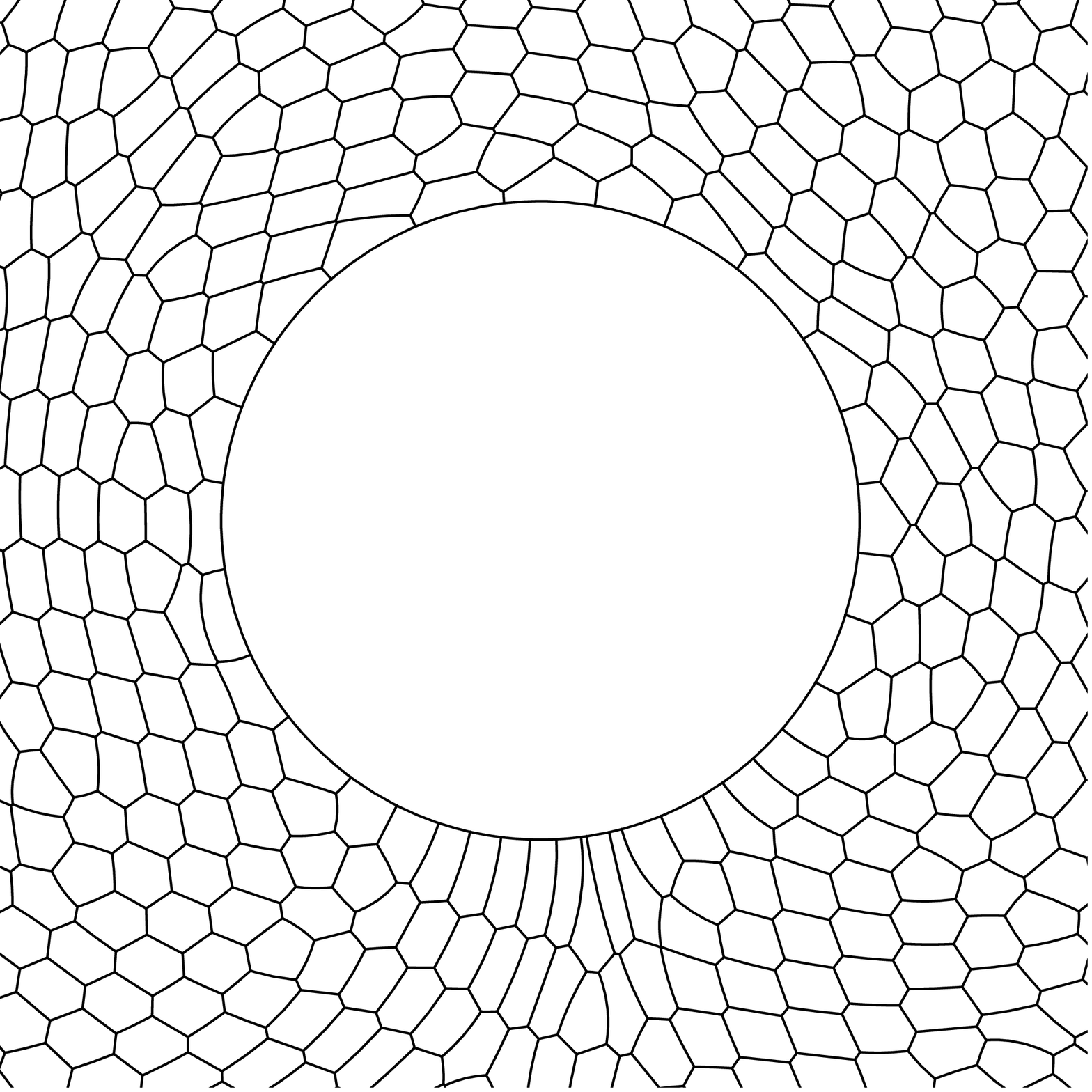}
{\bf(b)} }
\caption{
  Surface Evolver simulation of a very dry foam: zoom around the obstacle.
(a)   $\Phi = 3.7 \;  10^{-3}$;
(b)  $\Phi = 6  \;  10^{-6}$.
Here
      $d_0 = 4.8$ cm, $A = 0.16$  cm$^2$, $w_c = 10$ cm.
Flow from top to bottom. }
\label{Image_simul_dry}
\end{figure}

Surface Evolver simulations allow us to probe the range $10^{-5} < \Phi < 10^{-3}$. They indicate
that the force  saturates below $\Phi \sim 10^{-3}$, in agreement with our preliminary experiments of a foam confined between glass plates
(data not shown).

Direct observation of simulation images of very dry foams (Fig. \ref{Image_simul_dry}) confirms
that the up/downstream asymmetry in the number of bubbles touching the obstacle is
around 10, roughly independent of liquid fraction.

The model   seems to correctly describe the squashing and stretching
of bubble shapes.
However, the interpolation between both  extreme values
assumes a phenomenological expression
(eq. \ref{phenomeno}). It seems 
approximately valid only for $10^{-3}<\Phi< 10^{-1}$
 (see Appendix
\ref{Sec:Appendix_model}).
It applies to other obstacle shapes, such as an ellipse \cite{ellipse}.

\subsection{Yield drag \emph{versus} yield stress}

Princen and Kiss \cite{Princen1989}
have shown that in three dimensions a foam's yield stress scales as: $\sigma_Y = \gamma
(1-\Phi_{3D})^{1/3} Y(\Phi_{3D})/R_{32}$, where $R_{32}$ is the
surface-volume mean radius (Sauter radius), and $Y(\Phi_{3D})$ a decreasing function of
$\Phi_{3D}$ which is approximately $Y(\Phi_{3D}) \simeq -0.080 -
0.114\ln\Phi_{3D}$ \cite{Princen1989}.
 
At this stage, it is worth discussing the fundamental differences between
yield stress and yield drag. 

The yield stress or yield strain is an intrinsic property of the foam.
On the other hand, the yield drag depends on the geometry of the flow:
the foam does not yield \emph{everywhere}
around the obstacle, and especially not at angles $|\theta|
\approx \pi/4$ from the downstream direction, as appears both in
experiments (Fig. \ref{image_exp}) and in simulations (Figs.
\ref{Image_simul} and \ref{Image_simul_dry}).
This spatial dependence implies that the relation between
yield stress and yield drag is non-trivial and $\Phi$-dependent.
In particular, in a dryer foam (Fig. \ref{Image_simul_dry}),
the region where bubbles reach their maximal deformation is narrower \cite{dollet_local_en_preparation}.

Moreover, the flow around an obstacle involves not only shear, but also elongation,
especially near the front and back of the obstacle. When $\Phi$ decreases,
the bubble elongation
can become arbitrary large and dominates the contribution to the yield drag.

\section{Conclusion}

To summarise, we investigate the two-dimensional flow of a foam around a
circular obstacle, within a long channel. Our deterministic
(Surface Evolver) and stochastic (Potts model) simulations, as well as our
model and experiments, complement and validate each other.

The yield drag is defined as the low-velocity limit of the
interaction force between an obstacle and a flowing foam.
The network contribution scales as the
 ratio of obstacle to bubble diameter, as long as this ratio is larger than unity,
and is almost independent of the channel width. It increases
(because more and more stretched bubbles accumulate behind the obstacle)
  as a
power law  when the liquid fraction contained in the foam decreases
to $10^{-3}$,
then saturates.

Having found a relevant definition of the liquid fraction,
which is appropriate for experiments, simulations and theory, the
dependence of yield drag with liquid fraction is well
characterized. It is very different from that of local intrinsic
properties such as the yield stress or shear modulus.
   This observation  suggests  that
it will be difficult to deduce one quantity from the other. This
should be kept in mind in future simulations, and has to be taken
into account when modelling the foam behaviour.

Note that this definition of liquid fraction can be extended to other
2D flows in experiments (quasi-2D foam set-ups \cite{vazc05}) or simulations.
Extension to 3D
\cite{alonso,bruyn04} should also be simple,
especially since the main effect of quasi-2D set-ups -- external friction on
the glass plate \cite{irlandais,wang} --
   does not seem dominant   here.
Our present effective liquid fraction based on 
rheological properties (T1s) facilitates the comparison between 2D
and 3D flows 
(which is difficult when using the
 actual   volume fraction
of water 
\cite{denkov}). In simulations too, our definition immediately 
extends to 3D for both Surface Evolver, which uses as input parameter
the cut-off area for a face which undergoes a T1; and Potts model,
where the voxel (3D pixel) size plays the same role.

\section*{Acknowledgements}

We gratefully acknowledge the help of Steven Thomas with Potts
model simulations. We thank Isabelle Cantat for stimulating
discussions, and emphasising the differences between yield drag
and yield stress. CR thanks LANL and SC thanks LSP for
hospitality. YJ is supported by US DOE under contract
No. DE-AC52-06NA25396.SC is supported by EPSRC (EP/D071127/1) 
and his visit to Grenoble was supported by the Ulysses exchange programme.  Part of this work was performed during the FRIT workshop \cite{FRIT}.

\begin{appendix}

\section{Force measurements in experiments}
\label{Sec:force_meas_exp}

\subsection{Variation with foam thickness} \label{Sec:force_meas_exp_1}

\begin{figure}
\begin{center}
\includegraphics[width=8cm]{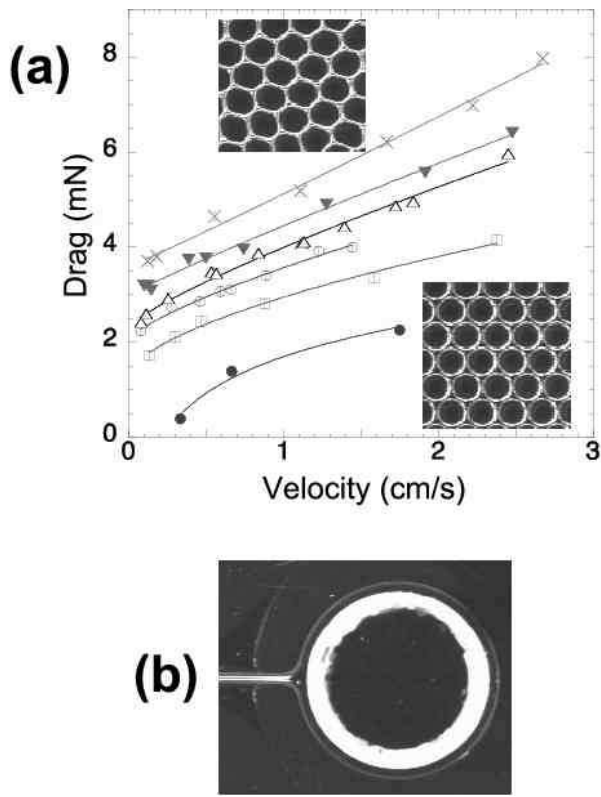}
\end{center}
\caption{(a) Drag \emph{versus} velocity for various distances $h$
between the top plate and the bottom solution: 2.0 ($\bullet$),
2.5 ($\square$), 3.0 ($\circ$), 3.5 ($\vartriangle$), 4.0
($\blacktriangledown$) and 4.5 mm ($\times$). Here
$A = 16.0$ mm$^2$ and $d_0 = 3$ cm.
The curves are the
best fits by the formula $F = F_Y^t + A\times V^a$. Snapshots of
the foam of the smallest (bottom) and highest (top) $h$ are also
displayed. (b) Photograph of a single soap film pulling on the
left side of the circular obstacle.} \label{Manip_force}
\end{figure}

We present here new data concerning the drag exerted by a flowing
foam of bubble area $A = 16.0$ mm$^2$ on a circular obstacle of
diameter $d_0 = 3$ cm. We measured the drag, as explained in full
detail in \cite{dolleteqrag05}, {\it versus} the foam velocity
$V$ for six different foam thicknesses (Fig. \ref{Manip_force}a).
As usual, the drag increases with increasing foam velocity. 

More
importantly for this paper, at given velocity, and especially at
the limit of vanishing velocity, the drag increases with
increasing foam thickness. This is due to (i) the decrease of liquid
fraction with increasing foam thickness, as shown by the
snapshots of the two extreme foam thicknesses in Fig.
(\ref{Manip_force}a); (ii) the increase in the height of the films with increasing foam thickness. 
We fit the data by the formula $F = F_Y^t + A V^a$ to get
the values of the total yield drag $F_Y^t$ for the various foam
thicknesses.

Notably, for the smallest foam thickness ($h=2.0$ mm) the
foam is almost decompacted and the drag tends to vanish at low
velocity. More precisely, the fit gives an unphysical negative
value. This suggests that the
rigidity loss transition \cite{Bolton1990} occurs for a foam thickness
between 2.0 and 2.5 mm.


\subsection{Comparison of network and total yield
drags} 
\label{Sec:force_meas_exp_2}

We measure the line tension directly as the force exerted on the
obstacle by a single soap film, as shown in Fig.
(\ref{Manip_force}b), for two thicknesses. Its value is 0.44 mN for
$h=4.0$ mm, and 0.49 mN for $h=4.5$ mm. 
This suggests that $\lambda /h \approx
110$ mN/m.

\begin{figure}
\begin{center}
\includegraphics[width=6cm]{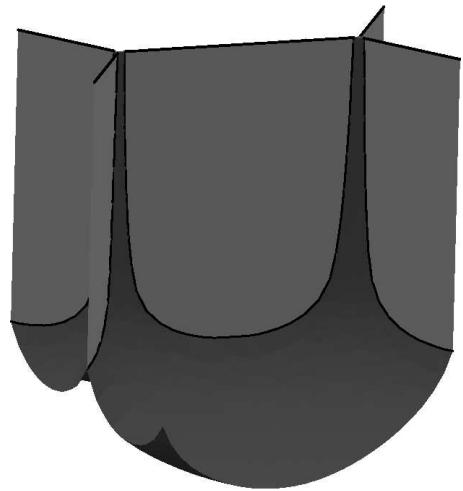}
\caption{\label{P(A)}
Shape of the interfaces between bubbles, for bubbles of area 16.0 mm$^2$
and volume $\mathcal{V} = 16.0\times 3.5$ mm$^3$, calculated
with the Surface Evolver. Vertical films are shown in light grey and the 
liquid surface in dark grey. To reproduce the experiment,
we enforce the hexagonal symmetry and include the buoyancy.
For simplicity, the junction
between lateral faces and the top plate is assumed to be orthogonal.
}
\end{center}
\end{figure}

The actual gas-liquid interfaces 
have a 3D curvature to match tangentially the water surface and the glass plate (Fig. \ref{P(A)}).
This explains why the the measured value is between a lower bound,
$\lambda /h= 2\gamma 
 \approx
52$ mN/m expected for a vertical soap film 
(that is, two flat gas/liquid interfaces),
and an upper bound, 
$\lambda /h = (2+\pi) \gamma  
 \approx
134$  mN/m, expected for two films with circular cross section
(see \cite{Dollet_these} for details).
 
Using the measured value of $\lambda$, 
we
determine the absolute value of
the network
yield drag $F_Y^n$ in experiments.
We check (data not shown)
 that 
$F_Y^n$ is consistently of the same order of magnitude, but lower than,
the value of $F_Y^t$  measured directly. The remaining part is attributed to the pressure
contribution $F_Y^p$, to be described in a further paper \cite{dollet_local_en_preparation}. The spatial variation of bubble height $h$ due to pressure differences is always less than 10\%, giving an upper limit to the spatial variations of $\lambda$.


\section{Variation of $F_Y^n$ with the cut-off length}
\label{Sec:Appendix_model}

We consider here only the bubbles touching the obstacle, and the contribution
of their walls  to the yield drag.
We assume that (i) the
foam is truly 2D and all bubbles have the same area (ideal 2D monodisperse foam); (ii) pressure
$P$ is the same for each bubble, i.e. bubble walls are straight and
all Plateau borders have the same radius of curvature  $R$; (iii) the obstacle
is   much
larger than the bubbles ($\sqrt{A} \ll d_0$) so that we can  neglect
its curvature at the
bubble scale. This latter approximation could in principle affect the
bubbles upstream, which share a long edge with the obstacle. However,
it should not greatly affect the bubbles downstream, which are the main
contributors to the drag.

\subsection{Geometry}

\begin{figure}
\begin{center}
\includegraphics[width=8cm]{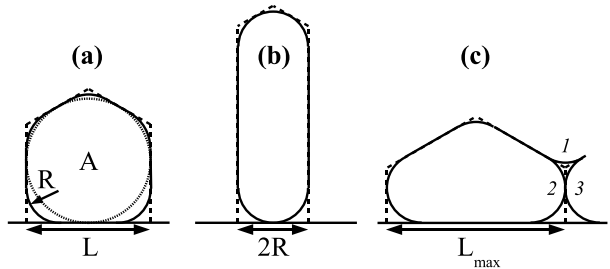}
\end{center}
\caption{Model configuration of the bubbles in contact with the
obstacle. (a) Equilibrium configuration: the dashed line
represents a polygonal bubble at $\Phi=0$, the dotted line is a
circular bubble at $\Phi =\Phi_c$, and the plain line
represents the intermediate case ($0<\Phi <\Phi_c$), with
straight edges and curved triangular vertices. (b) Configuration
at the limit of detachment: two neighbouring vertices on the
boundary of the obstacle come into contact. (c) Configuration at
the point of attachment of a new bubble. There is one vertex
between bubble 2, bubble 3 and the obstacle, and a second
vertex between bubbles 1, 2 and 3. When these two vertices  come in contact,
bubble 1 attaches to the obstacle.} \label{bubbles}
\end{figure}

    To model a wet foam, we
apply the decoration theorem \cite{Bolton1990}: the liquid is present only at
the vertices which decorate an ideally dry foam.
    For a bubble touching the obstacle, we denote by ${L}$
the distance between two neighbouring   vertices in contact with
the obstacle (Fig. \ref{bubbles}a).

When the foam flows, bubbles attach to the obstacle upstream, and
detach from it downstream.
Visual
observation of  both experiments (Fig. \ref{image_exp}) and
simulations (Fig. \ref{Image_simul}) indicate that bubbles are flattened
along the obstacle
at the leading side of the obstacle, and that they progressively stretch
streamwise at the trailing side.

${L}$ reaches its minimum value  downstream, where bubbles detach.
There, two
neighbouring  (decorated)  vertices come in
contact, and ${L}$ equals the  cut-off length $2R$ (Fig. \ref{bubbles}b).

On the other hand, for a new bubble to
attach to the obstacle upstream, two bubbles must detach through the configuration of
Fig.  (\ref{bubbles}c). In this case,  a
vertex between three bubbles merges with one between two bubbles and the wall.
The cut-off length is different, and now equals $(1+1/\sqrt{3})R$.
This
geometrically determines that the maximum bubble width ${L}_{\rm max}$ obeys:
\begin{equation}
A =
\left(1+\frac{1}{\sqrt{3}}\right)\; R{L}_{\rm max} +
\frac{{L}_{\rm max}^2}{4\sqrt{3}}.
\label{A(R)}
\end{equation}
Inverting eq. (\ref{A(R)}) yields  ${L}_{\rm max}$:
\begin{eqnarray}
{L}_{\rm max} (A,R)&=&
2  \sqrt{(\sqrt{3}+1)^2 R^2 + A\sqrt{3}} \nonumber \\
&& - 2  (\sqrt{3}+1)R .
\label{ell max}
\end{eqnarray}
At low liquid fraction,
${L}_{\rm max} $ tends to a finite value, namely $\sqrt{4A\sqrt{3} }$; there
is no singularity   at vanishing $R$.
Conversely, at high liquid fraction, ${L}_{\rm max} $ varies greatly
with $R$, so it is preferable to rewrite eq. (\ref{ell max}) and
determine $R$  from
the measurement of ${L}_{\rm max} $:
\begin{equation}
R (A,{L}_{\rm max} )=
\left(1+\frac{1}{\sqrt{3}}\right)^{-1}
\;
\left(\frac{A}{{L}_{\rm max} } - \frac{{L}_{\rm max}}{4\sqrt{3}}
\right).
\label{R(A)}
\end{equation}

\subsection{Continuous assumption}

   We assume that the shape of the bubbles
varies smoothly from the configuration of Fig.  (\ref{bubbles}c)
upstream to that of Fig.
(\ref{bubbles}b) downstream:
$2R < {L} < {L}_{\rm max}$.
Since the obstacle is much larger than the bubbles,
we switch from a discrete to a continuous description of the bubbles.
We thus consider ${L}$ as a continuous function
of the ortho-radial angle $\theta$ along the obstacle boundary:
$\theta=0$ downstream, $\pi$ (and $-\pi$) upstream.
Equivalently,
${L}^{-1}$ is the linear
density of vertices along the obstacle boundary.

Then ${L}(0)=2R$, ${L}(\pm \pi) = {L}_{\rm max}$.
To interpolate between these values, we assume
the following  phenomenological dependence,
reflecting that all bubbles  in the range $|\theta| \geq \pi/2$ appear squashed against the obstacle:
\begin{equation}
\begin{array}{ll}
|\theta|\leq\pi/2: & \quad {L}(\theta) = \displaystyle  \left(R+\frac{1}{2}L_{\rm max}\right) + \\
& \qquad \qquad  \displaystyle \left(R-\frac{1}{2}L_{\rm max}\right) \cos 2 \theta \\
|\theta|\geq\pi/2: & \quad {L}(\theta) = {L}_{\rm max}.
\end{array}
\label{phenomeno}
\end{equation}
Since each bubble edge exerts a pulling force
of magnitude $\lambda$ along  the outward normal vector of
the obstacle boundary, the network contribution to the drag is
\begin{equation}\label{FIntegral}
F = \frac{\lambda d_0 }{2} \; \int_{-\pi}^{\pi} \frac{\cos\theta}{{L}(\theta)}\;
\mathrm{d}\theta.
\end{equation}

To compute this integral, we introduce two dimensionless variables,
both functions of $A$ and $R$:
\begin{eqnarray}
\varepsilon &=& \frac{R }{ \sqrt{A}},\label{epsilon}\\
\beta &=&
\frac{{L}_{\rm max}}{2R}.
\label{beta}
\end{eqnarray}
The physical meaning of $\varepsilon$ is equivalent to
the liquid fraction, since
\begin{equation}
\Phi = (2\sqrt{3} -\pi) \varepsilon^2.
\label{Phi(epsilon)}
\end{equation}
On the other hand, $\beta$ quantifies the amount of up/downstream
asymmetry, that is, the squashing and stretching of bubbles. It
increases when $\Phi $ (or equivalently $\varepsilon$)
decreases (eq. \ref{ell max}):
\begin{equation}
\beta(\Phi) = \sqrt{ (\sqrt{3}+1)^2  +
\frac{(6-\sqrt{3} \pi)}{\Phi}      }    - (\sqrt{3}+1) .
\label{beta(epsilon)}
\end{equation}
When $\Phi$ goes to zero, $\varepsilon$ goes to zero too, and
$\beta$ diverges.

Using these variables, eq. (\ref{FIntegral}) yields
\begin{equation}
F = \frac{\lambda  \;  d_0}{{L}_{\rm max}}
\;
\left[  \;
\frac{ \beta}{\sqrt{\beta -1}} \arctan\left(\sqrt{\beta -1}\right) -
1 \right].
\label{F(phi)}
\end{equation}

At high liquid fraction, the force $F$ vanishes when $\beta = 1$, that is
(eq. \ref{beta(epsilon)}) when:
\begin{equation}
\Phi =  \frac{2\sqrt{3}-\pi}{2 + \sqrt{3}} = 0.086.
\end{equation}
At low liquid fraction, we develop eq. (\ref{F(phi)}) to leading
order in $\beta$ and insert the leading order term of eq.
(\ref{beta(epsilon)}) to obtain eq. (\ref{scaling_eq}).

\end{appendix}


\begin{thebibliography}{}
\bibitem{footnote1}
This is due to the left-right
symmetry of a bubble wall meeting the solid boundary.
This is independent of  the  hydrophilic or hydrophobic nature of the
solid surface (which  would only play a role, through the
presence or absence of a wetting film, in the energy dissipation,  not
relevant in the present paper).
\end{thebibliography}
\end{document}